\documentclass[aps,twocolumn,prb,showpacs,groupedaddress]{revtex4}
\usepackage{graphicx}
\begin{document}
\preprint{cond-mat/0204458}
\title{Self-trapped states and 
the related luminescence in PbCl$_2$ crystals}

\author{Masanobu Iwanaga}
\affiliation{Graduate School of Human and Environmental Studies, 
Kyoto University, Kyoto 606-8501, Japan}

\author{Masanobu Shirai}
\author{Koichiro Tanaka}
\affiliation{Department of Physics, Graduate School of Science, 
Kyoto University, Kyoto 606-8502, Japan}

\author{Tetsusuke Hayashi}
\affiliation{Faculty of Integrated Human Studies, Kyoto University, 
Kyoto 606-8501, Japan}

\date{\today}

\begin{abstract}
We have comprehensively investigated localized states of photoinduced 
electron-hole pairs with electron-spin-resonance technique and 
photoluminescence (PL) in a wide temperature range of 5--200 K. 
At low temperatures below 70 K, holes localize on Pb$^{2+}$ ions and 
form self-trapping hole centers of Pb$^{3+}$. The holes 
transfer to other trapping centers above 70 K. On the other hand, 
electrons localize on two Pb$^{2+}$ ions at higher than 50 K and form 
self-trapping electron centers of Pb$_2$$^{3+}$. 
From the thermal stability of the localized states and PL, we clarify that 
blue-green PL band at 2.50 eV is closely related to the self-trapped holes. 
\end{abstract}

\maketitle

\section{INTRODUCTION\label{intro}}
Electronic excited states relax into self-trapped (ST) states in solids 
where electrons strongly interact with acoustic phonons.~\cite{Toyozawa} 
The structure of ST states has been extensively examined with 
electron-spin-resonance (ESR) technique in ionic crystals.~\cite{Song,Silsbee}
The annihilation of the ST states often induces photons due to 
electron-hole ($e$-$h$) recombination via electric dipole transition. 
Therefore, luminescence study is another effective technique 
to investigate the ST states in high-efficient luminescent materials. 
Comprehensive study with ESR technique 
and luminescence spectroscopy is often powerful enough to clarify 
the ST states and the correlation with luminescence. 

\begin{figure}
\includegraphics[height=60.6mm,width=75mm]{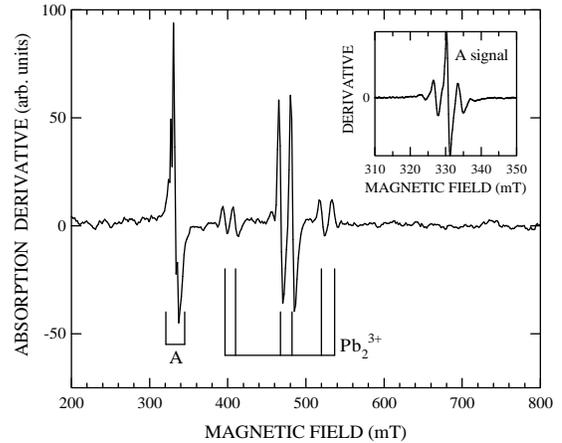}%
\caption{Typical ESR spectrum measured at 9 K after photoirrdiation at 80 K; 
the light source is described in Sec.\ \ref{experiment}. 
The inset enlarges ``A signal'' around 330 mT. 
Microwave frequency was 9.400 GHz. 
Magnetic field vector \textbf{B} was in $bc$ plane; the angle between 
the \textbf{B} and the $b$ axis was 10$^{\circ}$.}
\label{ESR80K}
\end{figure}

Localized states of excited electrons and holes in PbCl$_2$ crystals 
have been studied with 
ESR technique for the past few decades. In the early stage around 1970, 
the localized states induced by ultraviolet (UV) light irradiation at 80 K 
were observed around $g\approx 2$ (Refs.\ 
\onlinecite{Arends,Gruijter,Kerssen}). 
Figure \ref{ESR80K} shows ESR signals photoinduced at 80 K; 
a set of five resonances around $g\approx 2$ was named as 
``A signal''~\cite{Gruijter} and is enlarged in the inset. In addition, 
the ESR signals after x-ray irradiation at 10 K were also reported around 
$g\approx 2$, and were named as ``B signal'' and 
``C signal.''~\cite{Gruijter} 
Though some trials were performed,~\cite{Arends,Gruijter,Kerssen} 
the localized state responsible for the A signal 
has been disputable~\cite{Baranov} and the origins of the B and C 
signals have not been identified yet. 
The next advance was reported by Nistor \textit{et al}.~\cite{Nistor1} and 
Hirota \textit{et al}.~\cite{Hirota} 
in 1993; self-trapped electrons (STEL's) 
induced by x-ray~\cite{Nistor1} and $\gamma$-ray~\cite{Hirota} 
irradiation at 80 K were observed 
and found to be dimer-molecular Pb$_2$$^{3+}$ centers. 
The ESR spectrum of Pb$_2$$^{3+}$ is also presented in Fig.\ \ref{ESR80K}. 

Localized states in PbF$_2$ (Ref.\ \onlinecite{Pb3+2}) 
and PbBr$_2$ (Ref.\ \onlinecite{Iwanaga4}) have been revealed with ESR 
technique. In PbF$_2$, only holes get self-trapped and form Pb$^{3+}$ 
centers.~\cite{Pb3+2} On the other hand, both electrons and holes get 
self-trapped in PbBr$_2$, and form electron centers of Pb$_2$$^{3+}$ and 
hole centers of Br$_2$$^-$, respectively.~\cite{Iwanaga4} 
The configurations of self-trapped holes (STH's) are different 
between PbF$_2$ and PbBr$_2$. The difference is qualitatively explained by 
the components at the top of the valence bands; the $6s$ states of Pb$^{2+}$ 
are the main component in PbF$_2$, while the $4p$ states of Br$^-$ mainly 
constitute the valence band in PbBr$_2$ (Ref.\ \onlinecite{Fujita}). 
According to the cluster calculation~\cite{Fujita} for PbCl$_2$, 
the top of the valence band is composed of about half-and-half 
Pb$^{2+}$ ($6s$) and Cl$^-$ ($3p$). Therefore, the structure of localized 
hole centers in PbCl$_2$ is significant to clarify the relation of 
hole-relaxation dynamics with the electronic-band structure in PbCl$_2$ 
and to understand the relation in lead halides comprehensively. 

Luminescence in PbCl$_2$ has been studied in parallel to the ESR. 
The intense photoluminescence (PL) at low temperatures below 10 K 
was classified to UV-PL band at 3.8 eV, 
blue-PL band at 2.8 eV, and blue-green (BG) PL band at 2.5 
eV.~\cite{Gruijter,Gruijter2,Liidja} 
The UV-PL and the blue-PL bands are mainly induced 
under one-photon excitation into the exciton band, and the BG-PL band 
is dominantly induced under excitation into the energy region 
higher than the lowest exciton.~\footnote{Another PL band at 1.8--1.9 eV 
was reported. Because it is mainly induced under excitation into the 
low-energy range of the fundamental absorption including exciton absorption 
(Refs.\ \onlinecite{Gruijter,Gruijter2} and \onlinecite{Kitaura1}), 
the PL band is unlikely to be intrinsic PL. 
Therefore, the PL band is not discussed in this paper.}
As the origins of the UV-PL band at 3.76 eV and the blue-PL band at 2.88 eV, 
the self-trapped excitons (STE's) with the configuration of (Pb$^{+} +$ hole) 
were proposed, and the origin of the BG-PL band was attributed to the STE's 
of (Pb$_2$$^{3+} +$ hole).~\cite{Kitaura2,Kitaura1,Nistor4} 
The models claim that the lowest excitons and 
free $e$-$h$ pairs relax into different localized states. 
However, the Pb$^{+}$ centers in PbCl$_2$ have not been reported. 
Another interpretation on the PL bands was recently proposed;~\cite{Babin} 
it declares that the UV-PL and blue-PL bands come from the STE's of 
(Pb$_2$$^{3+} +$ hole) and the BG-PL band originates from the tunneling 
recombination of the pairs of a STEL of Pb$_2$$^{3+}$ and a STH of 
Cl$_2$$^-$. 
On the other hand, another PL study~\cite{Iwanaga3} has shown that the PL 
below 10 K under two-photon excitation into the exciton band is dominantly 
composed of the BG-PL band, and the intensity of the UV-PL band heavily 
saturates such that $I_{\rm UV}\propto I^{\,0.6}$ where $I$ is intensity of 
incident light. The results imply that, under the bulk excitation like 
two-photon excitation, the relaxed states of $e$-$h$ pairs are independent of 
the excitation energy and yield the BG-PL band. 
Consequently, it is possible enough to explain that the lowest excitons 
created under one-photon excitation are 
captured by surface defects and result in the UV-PL and the blue-PL bands. 
Thus, the origins of PL bands are still disputable. 

Excitons in PbBr$_2$ crystals, which belong to the same crystallographic 
group~\cite{Wyckoff} with PbCl$_2$ and have similar electronic-band 
structures,~\cite{Malysheva,Eijke,Kanbe,Fujita} 
undergo uncommon relaxation;~\cite{Iwanaga2} 
they spontaneously dissociate 
and relax into spatially-separated STEL's and STH's. 
The recent ESR study~\cite{Iwanaga4} supports the relaxed state of excitons 
structurally. A similar relaxation in PbCl$_2$ has been pointed 
out,~\cite{Iwanaga3} but the structures of the localized states 
have not been explored fully as already described. 

We have comprehensively examined the localized states of photoinduced 
$e$-$h$ pairs and PL properties in a wide temperature range of 5--200 K. 
As a result, the localized states of holes at low temperatures 
below 70 K have been found to form STH centers of Pb$^{3+}$; 
the structure of STH's is different from that inferred from the ESR 
spectrum at 80 K as shown in Fig.\ \ref{ESR80K} and from 
the recent study on PbBr$_2$ (Ref.\ \onlinecite{Iwanaga4}). 
We present the low-temperature ESR signals photoinduced below 10 K in 
Sec.\ \ref{ESRresults} and the PL properties in Sec.\ \ref{PLresults}. 
The STH's are analyzed with spin Hamiltonian in Sec.\ \ref{STH}. We 
discuss the correlation between ESR signals and PL in Sec.\ \ref{Thermal}, 
the origin of the A signal in Sec.\ \ref{OriginofA}, and 
the relaxation dynamics of $e$-$h$ pairs in comparison with PbBr$_2$ in 
Sec.\ \ref{dynamics}.

\section{EXPERIMENTS\label{experiment}}
Single crystal of PbCl$_2$ was grown with the Bridgman technique 
from 99.999\% powder. 
The crystal of orthorhombic D$_{2h}^{16}$ (Ref.\ \onlinecite{Wyckoff}) was 
cut in the size of 3$\times $3$\times $3 mm${^3}$ 
along the right-angled crystallographic $a, b$, and $c$ axes. 

The sample was photoirradiated with the second 
harmonics (pulsed 120-fs-width, 1-kHz, and 3.10-eV light) generated from 
a regeneratively amplified Ti:sapphire laser; average power of the 
incident light was about 20 mW$/$cm$^2$ on the sample surface. 
The incident photons induce two-photon interband transition, 
create $e$-$h$ pairs almost uniformly in the crystal, and produce 
measurable ESR signals within one minute. 
The samples were typically photoirradiated for five minutes. 
The irradiated sample was measured below 12 K with ordinary ESR technique 
in X-band range; resonant microwave frequency was 9.400$(\pm 0.004)$ GHz. 
The details of ESR and thermoluminescence (TL) measurements were previously 
reported in Ref.\ \onlinecite{Iwanaga4}. 
Raising- and lowering-rates of temperature in measuring TL were 0.5 K$/$s. 
In pulse-annealing measurement, the sample was kept for 
about one second at annealing temperature. 

\begin{figure}
\includegraphics[height=74.4mm,width=75mm]{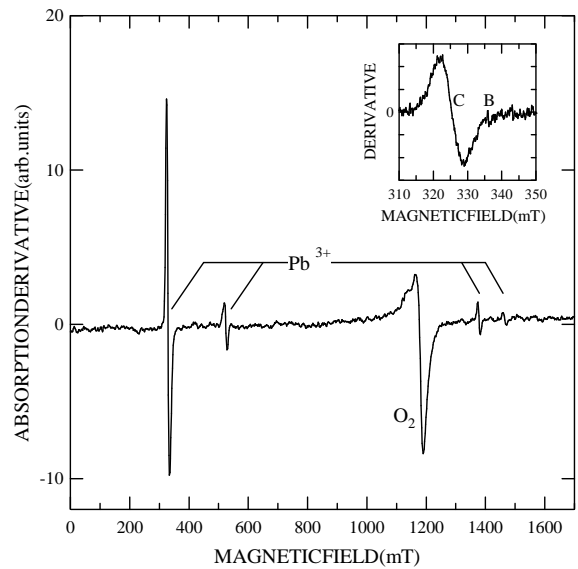}%
\caption{Typical ESR spectrum measured at 7 K after photoirrdiation at 6 K. 
The inset enlarges the signals around 330 mT. Microwave frequency was 
9.400 GHz. Magnetic field vector \textbf{B} was in the $bc$ plane; 
the angle between the \textbf{B} and the $b$ axis was 40$^{\circ}$.
Descriptions for Pb$^{3+}$, O$_2$, B, and C are given in 
Sec.\ \ref{ESRresults}.}
\label{ESR}
\end{figure}

PL was induced with the second harmonics (pulsed 5-ns-width, 10-Hz, 
and 2.33-eV light) generated from a Nd:YAG (yttrium aluminum garnet) laser; 
the incident light causes exciton-resonant two-photon 
excitation. The PL properties are essentially the same as those 
induced under one-photon excitation with UV light 
($\,\hbar\omega\ge$4.8 eV). 
The PL was detected with an intensified CCD camera 
with a grating monochromator, and time-resolved PL spectra were measured by 
gating the CCD camera; the temporal resolution was 5 ns.

\section{EXPERIMENTAL RESULTS}
\subsection{ESR Spectra and thermoluminescence\label{ESRresults}}

Figure \ref{ESR} shows typical ESR spectrum measured at 7 K 
after photoirradiation at 6 K; the ESR signals at 326, 522, 1400, 
and 1480 mT are readily induced together, while the signal at 1180 mT 
appears even in measuring empty capillary. Therefore, as reported in 
Ref.\ \onlinecite{Kon}, the signal at 1180 mT is ascribed to 
oxygen molecules in the capillary. 
The photoinduced four signals show similar thermal profile; besides, they 
saturate for microwave power higher than 
0.01 mW below 7 K, while keep linear response up to 0.1 mW at 10 K. 
Consequently, the four signals are ascribed to the same origin. 
The intense ESR signal at 326 mT, which is enlarged in the inset, 
consists of one broad resonance with 6-mT width; the structure 
differs from the fivefold split A signal in Fig.\ \ref{ESR80K}. 
The signals at 326 and 522 mT are almost independent of rotation angles. 
Though the two signals around 1400 mT depend on rotation angles, they 
degenerate when the magnetic field vector is parallel to the crystallographic 
axes. The reason for the split and the rotation-angle dependence is 
discussed in Sec.\ \ref{STH}. 
The intense line at 326 mT was induced by x-ray irradiation at 10 K 
and was named as C signal in Ref.\ \onlinecite{Gruijter}. 
The small signal at 336 mT in the inset of Fig.\ \ref{ESR} is the B signal 
named in Ref.\ \onlinecite{Gruijter}; it is far weaker than the C signal 
and hardly depends on rotation angles. From now on, we focus on the main 
signals of the A, the C, and the satellite signals. 

\begin{figure}
\includegraphics[height=69.9mm,width=75mm]{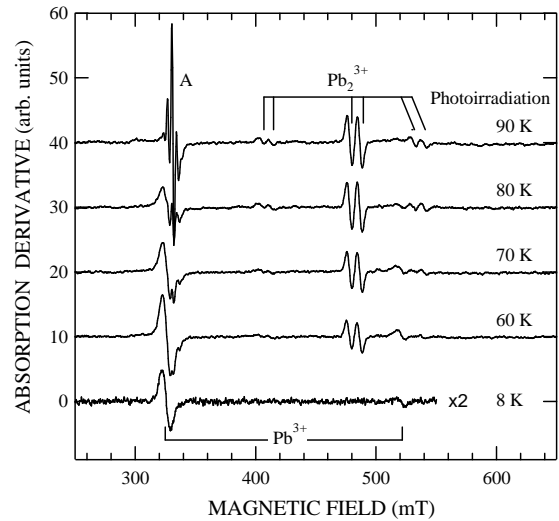}%
\caption{ESR spectra measured at 9 K after photoirradiation at various 
temperatures. They are displayed with the vertical offset. 
ESR spectrum at 8 K is enlarged by two times. Microwave 
frequency was 9.400 GHz. Magnetic field vector \textbf{B} was in the $bc$ 
plane; the angle between the \textbf{B} and the $b$ axis was 10$^{\circ}$.}
\label{ESR_irr}
\end{figure}

Except for 1180-mT resonance, the ESR spectrum in Fig.\ \ref{ESR} is 
very similar to that of Pb$^{3+}$ in KCl:Pb;~\cite{Pb3+1} 
therefore, the ESR signals in Fig.\ \ref{ESR} 
are ascribed to hole centers of Pb$^{3+}$. The hole centers are analyzed 
with spin Hamiltonian and are compared with Pb$^{3+}$ centers in other 
host crystals in Sec.\ \ref{STH}. 

Figure \ref{ESR_irr} displays the ESR spectra measured at 9 K 
after photoirradiation at 8, 60, 70, 80 and 90 K. 
A sequence of the spectra was measured by carrying out the 
photoirradiation and the ESR-data taking alternately. The ESR spectra 
after irradiation at 8--50 K are essentially the same in shape. At 
60 K, Pb$_2$$^{3+}$ signals become prominent, and the signals around 
330 mT are composed of Pb$^{3+}$ signals and other split signals at 60--80 K. 
The Pb$^{3+}$ signals are eventually replaced by the A signal at 90 K. 

\begin{figure}
\includegraphics[height=69mm,width=75mm]{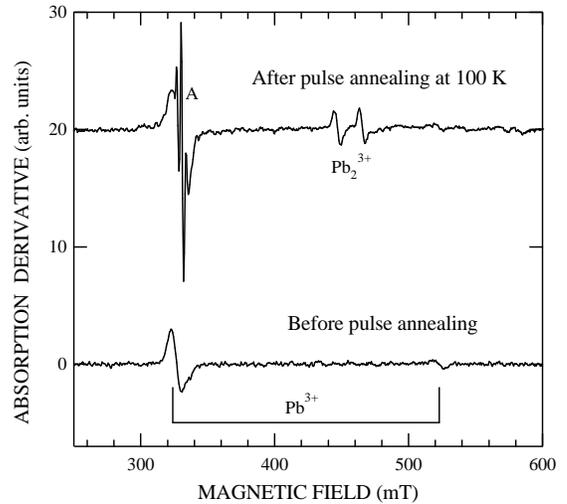}%
\caption{Photoinduced ESR signals before and after pulse annealing at 100 K. 
Both spectra were measured at 12 K. Microwave frequency was 9.400 GHz. 
Magnetic field vector \textbf{B} was the $bc$ plane; the angle between 
the \textbf{B} and the $b$ axis was 40$^{\circ}$.}
\label{ESR_ann}
\end{figure}

Figure \ref{ESR_ann} presents ESR spectrum (lower) after 
photoirradiation at 10 K, namely, before pulse annealing and ESR spectrum 
(upper) after pulse annealing at 100 K. Both spectra were measured at 12 K. 
The ESR spectrum before pulse annealing has the two prominent signals 
at 326 and 522 mT, which come from Pb$^{3+}$ centers. 
In the spectrum after pulse annealing, the signals from 
STEL centers of Pb$_2$$^{3+}$ appear around 460 mT, the A signal is 
observed around 330 mT, and the residual signal from Pb$^{3+}$ centers is 
still detected in the low-magnetic-field side of the A signal. 

\begin{figure}
\includegraphics[height=50.9mm,width=75mm]{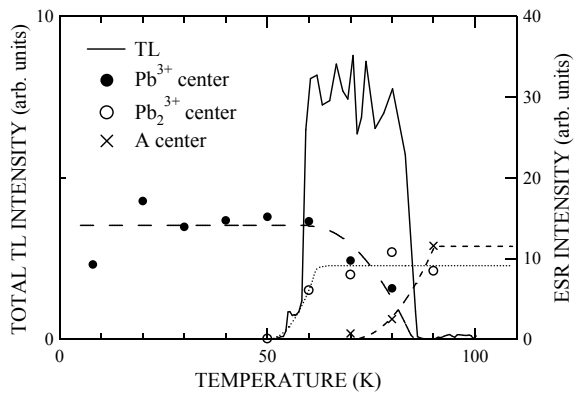}%
\caption{TL-growth curve (solid line) measured with raising temperature 
at 0.5 K$/$s. ESR intensity in Fig.\ \ref{ESR_irr} is shown for comparison; 
Pb$^{3+}$ (closed circle), Pb$_2$$^{3+}$ (open circle), and A centers 
(cross). Dashed, dotted, and broken lines are drawn for guides to the eye.}
\label{TLgrow}
\end{figure}

Figure \ref{TLgrow} shows TL-growth curve (solid line) and the intensity of 
ESR signals in Fig.\ \ref{ESR_irr} (open and closed circles and cross). 
TL is observed strongly at 60--85 K under the condition of raising 
temperature at 0.5 K$/$s. The TL spectrum in the temperature range 
is in agreement with the BG-PL spectrum. 

In a previous TL experiment,~\cite{Kitaura2} TL-growth curve measured with 
raising temperature at 0.1 K$/$s shows two discrete peaks at 51 and 74 K. 
The difference probably results from the temperature-rising rate; the 
rising rate in the previous report is slower than that in our measurement. 
The discrete peaks suggest the two different thermal activation in 50--80 K. 
The two thermal transfers were assigned to hole activation in 
Ref.\ \onlinecite{Kitaura2}. We discuss the activated carriers 
in Sec.\ \ref{Thermal}. 

\subsection{Photoluminescence\label{PLresults}}
\begin{figure}
\includegraphics[height=52.7mm,width=75mm]{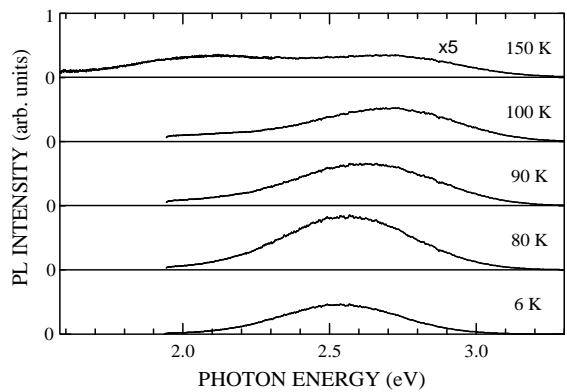}%
\caption{PL spectra at 6, 80, 90, 100, and 150 K under exciton-resonant 
two-photon excitation at 4.66 eV. PL spectrum at 150 K is enlarged by 
five times.}
\label{PL}
\end{figure}

Figure \ref{PL} presents PL spectra under exciton-resonant two-photon 
excitation at 4.66 eV. 
The PL spectra at 6--70 K are peaked at 2.52 eV and almost invariant in shape. 
Above 70 K, the PL spectrum begins to move toward high-energy side, 
and the peak reaches 2.7 eV at 100 K. 
In the range above 100 K, low-energy PL around 2 eV 
relatively increases with raising temperature, and finally the PL spectrum 
consists of broad PL bands at 150 K. 
The thermal behavior of PL induced under one-photon excitation into the 
fundamental absorption region ($\,\hbar\omega\ge 4.8$ eV) 
is the same as that in Fig.\ \ref{PL}. 

\begin{figure}
\includegraphics[height=62.2mm,width=75mm]{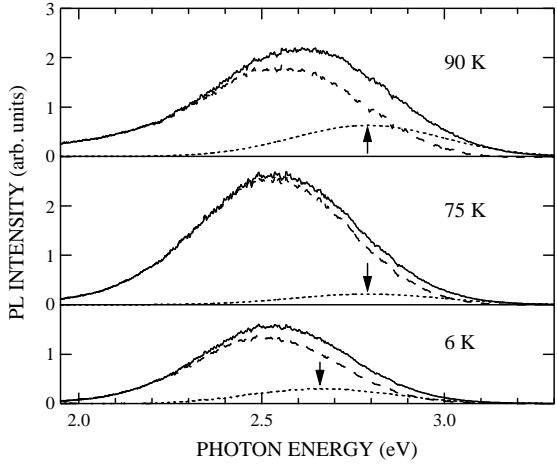}%
\caption{PL (solid line) and time-resolved PL (dashed and dotted lines) 
spectra at 6, 75 and 90 K under two-photon excitation at 4.66 eV. 
PL in $\mu$s range corresponds to blue PL (dotted line), and the rest of PL 
does to phosphorescent blue-green PL band (dashed line). 
Arrows indicate the blue-PL peaks, and are located at 2.66 eV 
for 6 K and at 2.79 eV for 75 and 90 K.}
\label{TRPL}
\end{figure}

Figure \ref{TRPL} shows PL spectra (solid line) and time-resolved PL (TRPL) 
spectra (dashed and dotted lines) under two-photon 
excitation at 4.66 eV, measured at 6, 75, and 90 K. 
The dashed line at 6 K represents the BG-PL band at 2.50 eV; 
the spectrum is obtained by measuring the PL over 1--99 ms, 
that is, by gating the CCD camera with 1-ms delay and 98-ms width. 
The BG-PL band decays phosphorescently; the intensity $I(t)$ is well 
described by $I(t)=(K/t)[1-\exp(-t/\tau)]$ where $K$ is a proportionality 
constant and $\tau = 100$ $\mu$s.~\cite{Iwanaga3}
The rest of the PL at 6 K is the PL band at 2.66 eV (arrow) 
and decays single-exponentially with 5.0 $\mu$s. This blue PL band is 
different from the blue PL band at 2.88 eV mentioned in Sec.\ \ref{intro}; 
the blue PL band at 2.88 eV is not induced under the present excitation. 
The PL spectra at 75 and 90 K are spectrally resolved with the TRPL 
spectra peaked at 2.79 eV (arrow). 
The blue PL band at 2.79 eV decays single-exponentially with 
2.6 $\mu$s at 75 K. Thus, the relative increase of the blue PL band results 
in the apparent high-energy shift of the PL spectrum in Fig.\ \ref{PL}. 

\begin{figure}
\includegraphics[height=56.7mm,width=75mm]{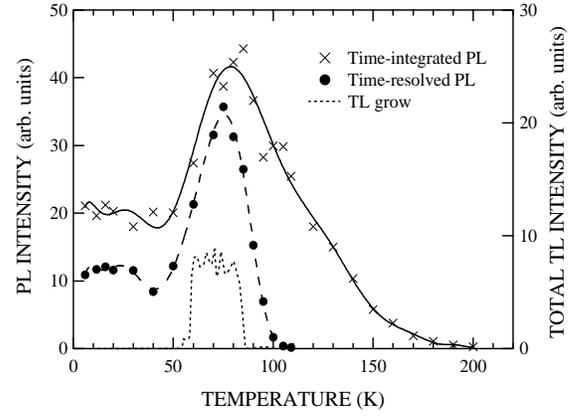}%
\caption{PL intensity integrated over 2.0--3.2 eV vs temperature; 
cross denotes time-integrated PL intensity over 0--99 ms and closed circle 
stands for time-resolved PL intensity over 1--99 ms. Solid and dashed lines 
are drawn with polynomial functions for guides to the eye. 
TL-growth curve (dotted line) is also presented for comparison.}
\label{PLvsT}
\end{figure}

PL intensity integrated over 2.0--3.2 eV at 6--200 K is plotted against 
temperature in Fig.\ \ref{PLvsT}; time-integrated PL intensity over 0--99 ms 
is represented with cross, and TRPL intensity over 1--99 ms with closed 
circle. PL intensity keeps almost constant below 50 K and increases in 
50--70 K. The increase is coincident with TL growth (dotted line), and the 
peaks of PL and TRPL intensity are located at about 80 K. 
Because the TRPL intensity corresponds to the phosphorescent component of 
PL, the increase of the PL intensity is ascribed to that of the 
phosphorescent BG-PL. Indeed, PL at 75 K decays in proportional to 
$t^{-1}$ for $t\ge 1$ ms. The TRPL intensity decreases in 80--100 K and is 
hardly observed above 100 K. The TRPL quenching is also coincident with the 
high-energy shift due to the relative increase of the blue PL band at 2.79 
eV as shown in Fig.\ \ref{TRPL}. Thus, the phosphorescent BG-PL band is 
induced below 100 K. Above 100 K, the blue PL band and other PL band around 
2 eV are dominantly induced and are finally quenched around 200 K. 

\section{DISCUSSION}
\subsection{Spin-Hamiltonian analysis of self-trapping hole centers%
\label{STH}}
The ESR signals at 326, 522, and about 1400 mT in Fig.\ \ref{ESR} 
show the same temperature profile and the intensity ratio 
$(326 \textrm{ mT}):(522 \textrm{ mT} + 1400 \textrm{ mT}) = 8:2$. 
The ratio is consistent with the isotope ratio of Pb ions; 
they have two isotope series such as $I=0$ and $1/2$, and the natural ratio 
is about $8:2$. If the intense signal at 326 mT corresponds to Zeeman 
transition, the $g$ value is estimated to be 2.06; the value larger than 
the free-electron $g$ value of 2.0023 implies that 
the center is a hole center.~\cite{Slichter} 
Indeed, from the comparison with Pb$^{3+}$ in KCl:Pb,~\cite{Pb3+1} 
the ESR signals are ascribed to hole centers of Pb$^{3+}$. 
The spin Hamiltonian $\mathcal{H}$ of the hole center is given by 
\begin{equation}
\mathcal{H}=\mu_B\mathbf{B}\cdot\underline{g}\cdot\mathbf{S} + %
\mathbf{S}\cdot\underline{A}\cdot\mathbf{I} - %
\mu_n\mathbf{B}\cdot\underline{g}_N\cdot\mathbf{I}, \label{STH_Ham}
\end{equation}
where $\mu_B$ denotes the Bohr magneton, $\mathbf{B}$ magnetic field vector, 
$\underline{g}$ the Zeeman tensor, $\mathbf{S}$ the electron spin, 
$\underline{A}$ the hyperfine tensor, $\mu_n$ the nuclear magneton 
of ${}^{207}$Pb 
($\mu_n = 0.5892\mu_N$, $\mu_N\textrm{: the nuclear magneton}$), 
$\mathbf{I}$ the nuclear spin, 
and $\underline{g}_N$ the nuclear Zeeman tensor. We note that the 
first-order hyperfine interaction of the hole with surrounding Cl-nuclei 
is not included in the Hamiltonian because our experimental results 
give no evidence of the superhyperfine (SHF) effect that the interaction 
splits each resonance at 326, 522, and 1400 mT in 
Fig.\ \ref{ESR} into tens of fine resonances as observed in 
PbF$_2$ (Ref.\ \onlinecite{Pb3+2}). Equation (\ref{STH_Ham}) describes 
the center that a hole strongly localizes on a Pb$^{2+}$ ion, namely, 
self-trapping hole center of Pb$^{3+}$. 

Equation (\ref{STH_Ham}) for $I=0$ is reduced to the electron 
Zeeman term. The intense signal at 326 mT in Fig.\ \ref{ESR} slightly 
depends on rotation angles. Spin Hamiltonian analysis for the Zeeman 
transition provides principal $g$ values as shown in Table \ref{STHpara}; 
we set $x=c$, $y=b$, and $z=a$ in this analysis. 

\begin{table}
\caption{Principal $g$ and $A$ values of spin Hamiltonian 
[Eq.\ (\ref{STH_Ham})]. The parameters of Pb$^{3+}$ centers in other host 
crystals are cited for comparison. The $A$ values are represented in GHz.}
\label{STHpara}
\begin{ruledtabular}
\begin{tabular}{c|cccc}
Host of Pb$^{3+}$ & & $g$ values & & $A$ values \\
 & $g_x$ & $g_y$ & $g_z$ & (GHz) \\
\colrule
PbCl$_2$\footnote{This work.} & 2.044 & 2.062 & 2.044 & 26.7 \\
& $\pm 0.001$ & $\pm 0.001$ & $\pm 0.001$ & $\pm 0.4$ \\
\colrule
KCl:Pb\footnote{Reference \onlinecite{Pb3+1}; $g$ tensor is 
assumed to be isotropic.} & & $2.034\pm 0.001$ & & $33.0 \pm 0.1$ \\
\colrule
ThO$_2$:Pb\footnote{References \onlinecite{Rohrig} and \onlinecite{Kolopus}.} 
& & $1.967 \pm 0.001$ & & $36.8 \pm 0.2$ \\
\colrule
ZeSe:Pb\footnote{Reference \onlinecite{Suto1}.} & & $2.072 \pm 0.001$ 
& & $20.7 \pm 0.1$ \\
\colrule
ZeTe:Pb\footnote{Reference \onlinecite{Suto2}; the accuracy of $g$ and $A$ 
velues is not reported.} & & 2.167 & & 15.7 \\
\end{tabular}
\end{ruledtabular}
\end{table}

\begin{figure}
\includegraphics[height=43.2mm,width=75mm]{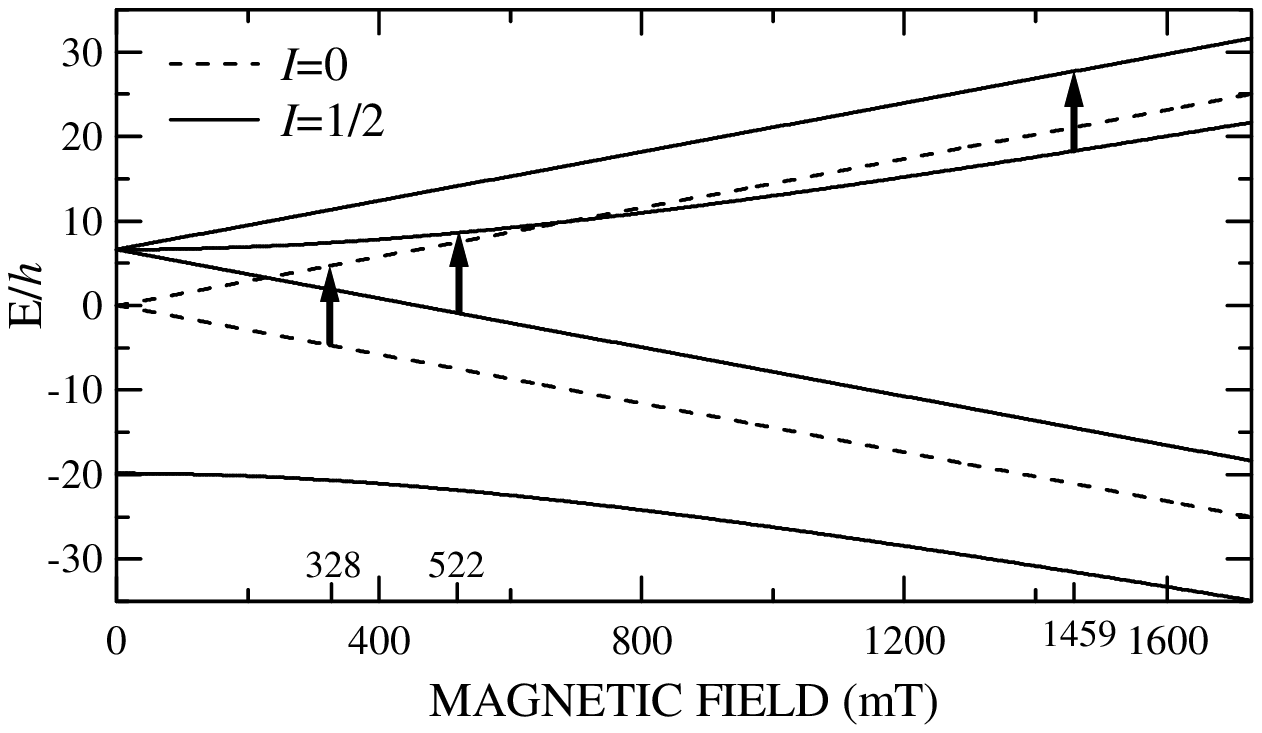}%
\caption{Energy diagram of spin Hamiltonian [Eq.\ (\ref{STH_Ham})] for 
$\textbf{B}\parallel a$. Broken lines: energy levels for $I=0$, describing 
electron Zeeman levels. Solid lines: energy levels for $I=1/2$. 
Arrows stand for observed ESR transitions at 9.404 GHz. 
The ordinate is represented in GHz.}
\label{E_diagram}
\end{figure}

In analyzing the spin Hamiltonian for $I=1/2$, we can assume that the $A$ 
tensor is isotropic because the resonance at 522 mT is independent of 
rotation angles. Furthermore, we replace, for simplicity, the contribution 
of nuclear Zeeman term to the energy eigenvalues with the effective nuclear 
$g_N$ value along the corresponding magnetic field. 
The simplification is justified because the contribution of the 
nuclear Zeeman term to Eq.\ (\ref{STH_Ham}) is small enough; 
indeed, the resonances are described by the electron Zeeman and isotropic 
hyperfine terms in a good approximation. Pb$^{3+}$ centers in other host 
crystals~\cite{Pb3+1,Rohrig,Kolopus,Suto1,Suto2} were analyzed within 
this approximation. However, to estimate $g_N$ values, 
we choose to include the effective value into the energy eigenvalues. 
The obtained equations of the eigenvalues are the same as Breit-Rabi 
formula,~\cite{Breit} which was applied to Pb$^{3+}$ centers in other 
hosts.~\cite{Pb3+1,Rohrig,Kolopus,Suto1,Suto2} Figure \ref{E_diagram} 
depicts energy diagram of Eq.\ (\ref{STH_Ham}) for $\textbf{B}\parallel a$; 
broken lines denote energy levels for $I=0$, 
and solid lines for $I=1/2$. Arrows in Fig.\ \ref{E_diagram} represent 
the ESR transitions observed in the measurement. Numerical estimation on 
$g_N$ gives the values between 0.1 and 6 for various directions of 
\textbf{B}. The anisotropy of the $g_N$ tensor is responsible for 
the rotation-angle dependence of the ESR signals around 1400 mT; 
moreover, the rotation-angle dependence indicates the deviation of principal 
$g_N$ axes from the $a$, $b$, and $c$ axes. 
In the unit cell of PbCl$_2$ crystal, there exist two-equivalent Pb-ion 
sites which reflect the crystallographic symmetry. 
The twofold splits around 1400 mT in Fig.\ \ref{ESR} are explained by the 
two possible configurations of Pb$^{3+}$ centers in the unit cell and by 
the anisotropy of the $g_N$ tensor. 

The evaluated $g$ and $A$ values are listed in Table \ref{STHpara} together
with the $g$ and $A$ values of Pb$^{3+}$ centers in other hosts. 
The $g$ values are about 2 in all hosts, while 
the $A$ values vary in 16--37 GHz. The $A$ values imply that the 
spatial spread of the hole is quite different in each host. 

\subsection{Thermal stability of ESR signals and PL bands\label{Thermal}}
Below 70 K holes localize on Pb$^{2+}$ ions and form 
self-trapping hole centers of Pb$^{3+}$, and, on the other hand, 
electron centers are not detected at 0--1700 mT below 50 K. 
As STEL centers with simple configuration, 
monomer Pb$^{+}$ and dimer Pb$_2$$^{3+}$ centers are possible in PbCl$_2$. 
Indeed, Pb$_2$$^{3+}$ centers are observed above 60 K. 
Pb$^{+}$ centers usually have $g$ value of 1.0--1.6 in Pb-doped alkali 
chloride;~\cite{Goovaerts2} in our measurement, 
the ESR signal would appear at 400--700 mT if it exists. However, 
the STEL centers of Pb$^{+}$ have not been detected. Thus, it is 
improbable that the Pb$^{+}$ centers are induced in PbCl$_2$. 

The sharp TL growth at 55 K in Fig.\ \ref{TLgrow} indicates 
the thermal activation of trapped carriers. 
Because the STEL centers of Pb$_2$$^{3+}$ appear at 60 K and the hole 
centers of Pb$^{3+}$ are quite stable at 60 K as shown in 
Fig.\ \ref{ESR_irr}, the activated carriers at 55 K are ascribed to 
electrons. However, the trapped state of electrons below 50 K has not been 
observed at 0--1700 mT. The electron center is most likely to has either of 
the following structures; (i) the center with ESR in the region higher than 
1700 mT, or (ii) the center without ESR. We discuss in Sec.\ \ref{dynamics} 
the relaxation dynamics of electrons below 50 K, producing the trapped 
state. 

The ESR spectra in Fig.\ \ref{ESR_irr} show that Pb$^{3+}$ centers thermally 
change into the A center above 70 K and are finally replaced at 90 K. 
In a pulse-annealing experiment~\cite{Nistor2} on PbCl:Tl, the A signal 
is weaken under annealing at 200 K, and the ESR signals coming from 
hole-trapping centers of Tl$^{2+}$ grow at the temperature. The result 
implies that the A signal originates from hole-trapping centers.
The STEL centers of Pb$_2$$^{3+}$ appear 
above 60 K as shown in Figs.\ \ref{ESR_irr} and \ref{ESR_ann}, 
become unstable thermally above 130 K, and are quenched 
around 210 K.~\cite{Nistor1,Nistor2} 

\begin{figure}
\includegraphics[height=44mm,width=66mm]{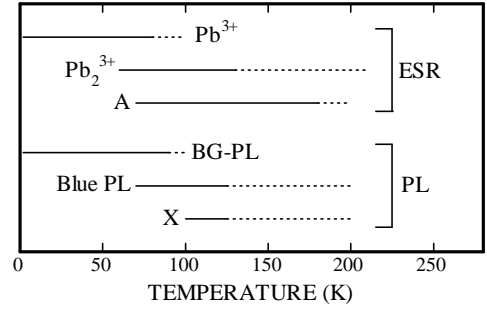}%
\caption{Schematic diagram for thermal stability of the photoinduced ESR 
signals and PL bands. 
Solid line: the stable range of the signals. Dotted line: 
the unstable but still detectable range. Blue PL denotes the blue PL band at 
2.79 eV in Fig.\ \ref{TRPL}. X stands for the broad PL band around 2.1 eV 
in Fig.\ \ref{PL}}
\label{ESRvsPL}
\end{figure}

Figure \ref{ESRvsPL} summarizes and schematically represents 
the thermal stability of ESR signals and PL bands; solid lines stand for 
the stable range, and dotted lines denote unstable but still detectable 
range. Blue PL in Fig.\ \ref{ESRvsPL} denotes the blue PL band at 2.79 eV 
in Fig.\ \ref{TRPL}, and X does the broad PL band around 2 eV in 
Fig.\ \ref{PL}. Obviously, the BG-PL band is thermally coincident with 
the STH centers of Pb$^{3+}$. In addition, the thermal activation of 
electrons at 55 K corresponds to the TL which has the same shape with 
the BG-PL band. Consequently, we ascribe the luminescent 
state yielding the BG-PL band to the STE associated with 
a Pb$^{3+}$ center; probably, the STE has the configuration of 
(Pb$^{3+}$ $+$ electron). The disagreement of the BG-PL band with 
Pb$_2$$^{3+}$ centers in Fig.\ \ref{ESRvsPL} excludes the model for the 
BG-PL band proposed in Refs.\ 
\onlinecite{Kitaura1,Kitaura2,Nistor4,Babin}.\footnote{From the experimental 
evidence presented already, we can also discuss the models for the UV-PL and 
2.88-eV blue-PL bands in Refs.\ \onlinecite{Kitaura1,Kitaura2,Nistor4,Babin}; 
the absence of Pb$^+$ centers denies the model in 
Refs.\ \onlinecite{Kitaura1,Kitaura2,Nistor4}, and the disagreement of the 
temperature range of Pb$_2$$^{3+}$ centers with those of the UV-PL and 
blue-PL bands excludes the model in Ref.\ \onlinecite{Babin}.} 
The blue PL band at 2.79 eV almost corresponds to the STEL centers of 
Pb$_2$$^{3+}$ and the A centers. 
Therefore, this study provides a possibility that the blue PL band 
comes from the STE's associated with Pb$_2$$^{3+}$ centers, but, to identify 
the origin, the further study with optically detected magnetic resonance 
technique is surely needed. 

\subsection{Origin of ``A signal''\label{OriginofA}}
The origin of the A signal has been disputed over the past 
three decades; it was once attributed to electron centers of Pb$^{+}$ 
(Refs.\ \onlinecite{Gruijter} and \onlinecite{Kerssen}), 
but the properties of the A signal, the $g$ value and the ESR spectrum, 
are far different from those of Pb$^{+}$ in other host 
crystals.~\cite{Baranov} 

The A signal was recently assigned to self-trapping hole centers 
of Cl$_2$$^{-}$ (V$_{\rm K}$ centers).~\cite{Nistor2} 
The A signal is composed of five resonances as presented in 
Fig.\ \ref{ESR80K}; the intensity ratio at 9 K is estimated to be 
$1:10:27:10:0.5$. 
However, the ESR spectrum of Cl$_2$$^{-}$ splits into seven resonances in 
the first-order hyperfine effect, and the intensity ratio of $1:2:3:4:3:2:1$ 
disagrees with that of the A signal. 
Moreover, the rotation-angle dependence of the A signal~\cite{Gruijter} 
does not show the anisotropy peculiar to the dimer-molecular V$_{\rm K}$ 
centers,~\cite{Silsbee} but is almost isotropic. 
Thus, the A signal is unlikely to originate from the 
V$_{\rm K}$-type hole centers of Cl$_2$$^{-}$. 

Furthermore, the A signal of five resonances does not come from the 
Pb$^{3+}$ centers with SHF interaction as observed 
in PbF$_2$ (Ref.\ \onlinecite{Pb3+2}), 
because (i) the satellite at 522 mT disappears up to 90 K as shown 
in Fig.\ \ref{ESR_irr} and (ii) the SHF structure, which stems from the 
interaction of the Pb$^{3+}$ centers with the surrounding Cl$^{-}$ ions, 
is composed of tens of fine resonances.~\cite{Pb3+1} 

As shown in Fig.\ \ref{ESR_ann}, the A signal appears under pulse annealing 
at 100 K after photoirradiated at 10 K. This result indicates 
that the localized state responsible for the A signal is produced by thermal 
transfer of the localized states induced at 10 K. 
In addition, Fig.\ \ref{ESR_irr} shows that the STEL centers of 
Pb$_2$$^{3+}$ are stable over 80--90 K where the Pb$^{3+}$ signals 
change into the A signal. Consequently, we declare that the A signal is 
formed by either of the two following ways: (i) The STH centers become 
unstable around 80 K and the holes transfer to other trapped state 
associated with permanent lattice defects. (ii) Vacancies begin to move 
thermally around 70 K, affect the STH centers, and modify 
the STH centers or make the STH center unstable. 
In any way, the A signal is ascribed to the hole-trapping centers associated 
with the permanent lattice defects such as vacancies or impurities or both. 

\subsection{Relaxation dynamics of electron-hole pairs\label{dynamics}}
We discuss here the relaxation of $e$-$h$ pairs in PbCl$_2$ from the 
comparison with relaxation in PbBr$_2$ (Ref.\ \onlinecite{Iwanaga2}).

In PbBr$_2$, the spontaneous dissociation of $e$-$h$ pairs was 
reported,~\cite{Iwanaga2} 
and the individual self-trapped states of both electrons and holes 
have been recently evidenced with ESR technique.~\cite{Iwanaga4} 

However, the present ESR study on PbCl$_2$ shows that 
the relaxed states of $e$-$h$ pairs are STH's and trapped 
electrons below 50 K, STH's and STEL's at 60--70 K, and the A centers 
and STEL's at 80--130 K (Fig.\ \ref{ESRvsPL}). 
The complicated results probably come from the inevitably dense vacancies in 
PbCl$_2$;~\cite{Verwey2,Ober} PbCl$_2$ crystals are high-ionic conductors 
and therefore include dense anion vacancies more than 10$^{17}$ cm$^{-3}$. 
In the crystals, the anion vacancies or the vacancy-associated defects are 
electron traps and can be competitors to the STEL centers of 
Pb$_2$$^{3+}$. Indeed, the thermal production of the STEL's in 
Fig.\ \ref{ESR_ann} suggests that the competitors are efficient enough to 
result in the absence of STEL's below 50 K. 

According to the theoretical study by Sumi,~\cite{Sumi} 
the exciton--acoustic-phonon interaction determines the relaxed state of 
$e$-$h$ pairs and classifies the relaxed states by the strength and the sign 
of the coupling constants. 
In the theoretical study, the crystallographic field is 
idealized with omitting the lattice defects such as dense vacancy, 
so that the study is not applicable to PbCl$_2$ straightforwardly. 
However, the coexistence of STEL's and STH's at 60--70 K shows the evidence 
that both electrons and holes strongly interact with acoustic phonons. 
Consequently, though the localized states are far more complicated 
in the real crystal, we believe that the $e$-$h$ relaxation similar to that 
in PbBr$_2$ does also realize in PbCl$_2$. 

At the end of discussion, we compare the relaxed states of $e$-$h$ pairs 
in PbF$_2$, PbCl$_2$, and PbBr$_2$. Though the crystallographic structure of 
$\beta$-PbF$_2$ is cubic and those of PbCl$_2$ and PbBr$_2$ 
are orthorhombic,~\cite{Wyckoff} STH centers in $\beta$-PbF$_2$ (Ref.\ 
\onlinecite{Pb3+2}) and PbCl$_2$ are monomer Pb$^{3+}$ centers while STH 
centers in PbBr$_2$ are dimer Br$_2$$^-$ centers.~\cite{Iwanaga4} 
The relaxed states indicate that holes in $\beta$-PbF$_2$ and PbCl$_2$ 
strongly interact with Pb$^{2+}$ ions while holes in PbBr$_2$ do with Br$^-$ 
ions. The top of the valence band in PbCl$_2$ is composed of 
about half-to-half Pb$^{2+}$ ($6s$) and Cl$^-$ ($3p$),~\cite{Fujita} but the 
holes nevertheless localize only on Pb$^{2+}$ ions. 
On the other hand, though the bottoms of the conduction band in 
$\beta$-PbF$_2$, PbCl$_2$, and PbBr$_2$ are composed of the $6p$ states of 
Pb$^{2+}$ ions,~\cite{Nizam,Fujita} electrons in $\beta$-PbF$_2$ do not 
get self-trapped while electrons in PbCl$_2$ and PbBr$_2$ form STEL centers 
of Pb$_2$$^{3+}$. From the comparisons here, we remark that in lead halides 
the structure of electronic band does not simply determine the relaxed states 
of $e$-$h$ pairs, but the electron-phonon and the hole-phonon interaction 
plays a crucial role to determine the relaxed states.

\section{CONCLUSIONS}
We have comprehensively investigated photoinduced ESR signals and 
PL in a wide range of 5--200 K in PbCl$_2$. 
As a result, the hole centers below 70 K have been found to be 
self-trapping centers of Pb$^{3+}$. The ESR study in the wide temperature 
range reveals the thermal change of localized centers; 
in particular, the grow of the STEL centers of Pb$_2$$^{3+}$ and the 
thermal transfer from the STH centers to the A centers. 
From the comparison with the thermal stability of the localized centers, 
the origins of PL bands have been discussed; 
we finally conclude that the STH centers of Pb$^{3+}$ are responsible 
for the STE's yielding the BG-PL band at 2.50 eV.

\begin{acknowledgments}
We would like to thank Mr.\ I.~Katayama and Mr.\ T.~Hasegawa for technical 
assistance in ESR experiments. 
\end{acknowledgments}

\bibliography{iwanaga}

\end{document}